\documentclass[twocolumn,aps,superscriptaddress,showpacs,nofootinbib,floatfix]{revtex4-1}
\usepackage{epsfig,bm,feynmf}
\usepackage{graphics}
\usepackage{amsmath}
\usepackage[eps]{pstricks}
\usepackage{physics}
\usepackage{slashed}
\usepackage{dsfont}
\usepackage{multirow}
\usepackage{hyperref}
\usepackage{subfigure}
\usepackage[normalem]{ulem}  % \sout{old text} for strikeout

\renewcommand\sout{\bgroup \color{red} \ULdepth=-.5ex \ULset}
\begin{document}

%%%%%%%%%%%%%%%%%%%%% Title %%%%%%%%%%%%%%%%%%%%%%

\title{Dynamically generated axial-vector meson resonance in the chiral symmetry restored vacuum}

%%%%%%%%%%%%%%%%%%%% Authors %%%%%%%%%%%%%%%%%%%%%

\author{Jisu Kim}%
\email{fermion0514@yonsei.ac.kr}
\affiliation{Department of Physics and Institute of Physics and Applied Physics, Yonsei University, Seoul 03722, Korea}

\author{Su Houng Lee}%
\email{suhoung@yonsei.ac.kr}
\affiliation{Department of Physics and Institute of Physics and Applied Physics, Yonsei University, Seoul 03722, Korea}

%%%%%%%%%%%%%%%%%%%% Abstract %%%%%%%%%%%%%%%%%%%%%

\begin{abstract}

We study the modification of the properties of the axial-vector meson, dynamically generated through the unitarization procedure, in the vacuum where the chiral symmetry is restored. 
This is accomplished by scaling the pion decay constant as the chiral order parameter while keeping the other input parameters fixed.
We find that the mass and width of the axial-vector meson reduce to those of the vector meson, as anticipated by the Weinberg sum rules.
The findings are consistent with the results of a recent QCD sum rule calculation, wherein the chiral order parameter is expressed through chiral symmetry-breaking four-quark operators, leading to the mass-squared difference scaling in proportion to variations in the chiral order parameter.
We calculate the scaling behavior for the mass differences obtained from the unitarization method using both the physical and massless pion masses.
\end{abstract}

%\pacs{25.75.Nq, 25.75.Ld}
%\keywords{}

\maketitle

\section{Introduction}
Understanding the generation of  hadron mass from the underlying QCD dynamics is a long-standing fundamental problem.
Spontaneous chiral symmetry breaking is believed to play an important role in generating the masses of hadrons\cite{Hatsuda:1985eb,Brown:1991kk,Hatsuda:1991ez,Hatsuda:1994pi,Leupold:2009kz}.
According to the Weinberg sum rules \cite{Weinberg:1967kj}, chiral symmetry breaking is responsible for the mass difference between chiral partners, and their masses in the chiral symmetry-restored vacuum should be degenerate.
However, the Weinberg sum rule does not offer information on how the vector meson mass itself should depend on the chiral order parameter. Furthermore, it remains unclear how the mass differences between chiral partners scale with the chiral order parameter.

Recently, we evaluated the degenerate masses of chiral partners in the QCD sum rules approach by removing the chiral symmetry-breaking effects from the quark operators\cite{Kim:2020zae,Kim:2021xyp}. 
We found that the masses of chiral partners do indeed become degenerate, with these masses being slightly lower than those of the lighter particles among the chiral partners.

In this paper, we will investigate how the mass difference between chiral partners scales with the chiral order parameter. To achieve this, we will consider scenarios where the axial-vector mesons are dynamically generated through the unitarization procedure, followed by studying the effects of chiral symmetry restoration through alterations in the pion decay constant.

\section{Unitarization Procedure}
In Ref.~\cite{Roca:2005nm}, the low-lying axial-vector meson resonances were studied by calculating the unitarized scattering amplitudes generated from the $s$-wave interaction of vector mesons with pseudoscalar mesons. 
To study the effect of chiral symmetry restoration in the flavor SU(2) sector and to compare the results with those obtained previously by us using QCD sum rule methods\cite{Kim:2020zae,Kim:2021xyp}, we will begin our discussions on the axial-vector mesons generated through the $\rho$-$\pi$ channel, wherein the pions are considered massless. Then, to compare the behaviors of the chiral and parity partners, we consider the symmetry-broken SU(3) case with the physical input parameters given in Table~\ref{sr_parameter}.

The interaction of two-vector and two-pseudoscalar mesons at the lowest order can be obtained from the following interaction Lagrangian\cite{Birse:1996hd}.
\begin{equation} 
\begin{split} 
\mathcal{L}_{I} = - \frac{1}{4}\Tr[(\grad_{\mu}V_{\nu} - \grad_{\nu} V_{\mu})(\grad^{\mu}V^{\nu} - \grad^{\nu} V^{\mu})],
\end{split} 
\end{equation}
where $\Tr$ runs over 3 flavors. The covariant derivative $\grad_{\mu}$ and the vector current $\Gamma_{\mu}$ are, respectively, defined as
\begin{equation} 
\begin{split} 
\grad_{\mu}V_{\nu} =&\;\; \partial_{\mu}V_{\nu} + [\Gamma_{\mu},V_{\nu}],\\
\Gamma_{\mu} =&\;\; \frac{1}{2}(u^{\dagger}\partial_{\mu}u + u \partial u^{\dagger}),\\
u^{2} =& e^{i(\sqrt{2}/f)P}.
\end{split} 
\end{equation}
Here, $f$ is the pion decay constant, and $P(V)$ is the pseudoscalar meson octet(vector meson nonet) SU(3) matrix. Expanding the Lagrangian $\mathcal{L}_{I}$ up to two-pseudoscalar fields, one can obtain the two-vector-two-pseudoscalar interaction Lagrangian,
\begin{equation} 
\begin{split} 
\mathcal{L}_{VVPP} = -\frac{1}{4 f^{2}} \Tr([V^{\mu},\partial^{\nu} V_{\mu}][P,\partial_{\nu}P]),
\end{split} 
\end{equation}
accounting for the Weinberg-Tomozawa interaction for $VP \to VP$ process\cite{Lutz:2003fm,Weinberg:1966kf,Tomozawa:1966jm}.

Solving the Bethe-Salpeter equation with the interaction terms above, one can evaluate the transition matrix $T$ as follows. 

\begin{equation} 
\begin{split} 
T = \frac{-V}{1+V\hat{G}}\vec{\epsilon}\cdot \vec{\epsilon}\,',
\end{split} 
\end{equation}
where $\hat{G} = G(1+\frac{1}{3} \frac{q^{2}}{m_{\rho}^{2}})$,
\begin{equation} 
\begin{split} 
V(s) = \frac{\epsilon\cdot \epsilon'}{4f^{2}_{\pi}} \bigg[& 3s - 2(m^{2}_{\rho} +m^{2}_{\pi}) - \frac{1}{s}(m_{\rho}^{2} - m^{2}_{\pi})^{2}\bigg],
\label{int_lag}
\end{split} 
\end{equation}
\\
\begin{equation} 
\begin{split} 
G(\sqrt{s}) = &\frac{1}{16\pi^{2}} \bigg\{a(\mu) + \ln\frac{m^{2}_{\rho}}{\mu^{2}} + \frac{m_{\pi}^{2}-m^{2}_{\rho} + s}{2s}\ln\frac{m_{\pi}^{2}}{m_{\rho}^{2}}\\
&+\frac{q}{\sqrt{s}}[\ln(s - (m_{\rho}^{2}-m_{\pi}^{2}) +2q\sqrt{s}) \\
&+\ln(s+(m_{\rho}^{2} - m_{\pi}^{2})+2q\sqrt{s} )\\
&-\ln(s-(m_{\rho}^{2} - m_{\pi}^{2})-2q\sqrt{s} )\\
&-\ln(s+(m_{\rho}^{2} - m_{\pi}^{2})-2q\sqrt{s})-2\pi i]\bigg\},
\label{loopfn}
\end{split} 
\end{equation}
\begin{equation} 
\begin{split} 
q = \frac{1}{2\sqrt{s}} \sqrt{[s - (m_{\rho}+m_{\pi})^{2}][ s - (m_{\rho} -m_{\pi})^{2}]},
\label{com-momentum}
\end{split} 
\end{equation}
$\epsilon(\epsilon')$ and $m_{\rho}$($m_{\pi}$) stand for the polarization four-vector of incoming(outgoing) vector meson and the mass of $\rho$($\pi)$ meson, respectively.
In this paper, we take the loop integral $G(\sqrt{s})$ evaluated through dimensional regularization. 

As explained in Refs. \cite{PhysRev.107.1148, Chan:1961}, bound states appear as poles on the positive imaginary $q$-axis, whereas resonances appear as poles in the lower half-plane of the complex $q$ plane. Thus for $\Re(\sqrt{s})$ above the lowest threshold $m_{\rho}+ m_{\pi}$, the loop function of Eq.~\eqref{loopfn} is used with the solution for the square root in  Eq.~\eqref{com-momentum} with $\Im(q)<0$ (second Riemann sheet) to look for the pole position. Meanwhile, for  $\Re(\sqrt{s})$ below the lowest threshold, the loop function $G^{I}(\sqrt{s})$,
\begin{equation} 
\begin{split} 
G^{I}(\sqrt{s}) = G(\sqrt{s}) - i \frac{q}{4\pi \sqrt{s}},
\end{split} 
\end{equation}
is used with $\Im(q)>0$ and $\Re(q) = 0$ (first Riemann sheet). We chose the subtraction constant $a(\mu = 1\;\mathrm{GeV})$ to be -1.85 given as an optional choice in Ref.~\cite{Roca:2005nm}. For the resonance states, we employed the same value of $a(\mu)$ across varying values of the decay constant $f_{\pi}$.
For the bound states investigated in \ref{su3}, we chose the subtraction constant to be -4.08 to ensure the continuity of the $a_{1}$ meson mass shift. 
Also, note that once the subtraction constant is set, it does not depend on the scale $\mu$, as the shift of $\ln(\mu^2)$ in Eq.~\eqref{loopfn} is canceled through the change in $a(\mu)$.
\begin{equation} 
\begin{split} 
a'(\mu') - a(\mu) = 2\ln(\mu/\mu').
\end{split} 
\end{equation}

Following the Gell-Mann-Oakes-Renner (GMOR) relation, 
\begin{equation} 
\begin{split} 
f^{2}_{\pi} m_{\pi}^{2} = - (m_{u} + m_{d}) \ev{\bar{q}q},
\end{split} 
\end{equation}
we consider the square of the pion decay constant $f_{\pi}^2$ as the chiral order parameter, analogous to the quark condensates in the QCD sum rules approach\cite{Kim:2020zae, Kim:2021xyp}. 
As a consequence of the chiral symmetry breaking, the axial current couples to the pion state,
\begin{equation} 
\begin{split} 
\bra{0}\mathcal{J}^{a}_{\mu,5}(x)\ket{\pi^{b}(q)} = iq_{\mu}f_{\pi}\delta_{ab}e^{-iq\cdot x}.
\end{split} 
\end{equation}
This implies that the pion decay constant plays the role of the chiral order parameter at the hadron level.
The input values for parameters appearing in the calculations are given in Table \ref{sr_parameter}.

\begingroup
\begin{center}
\setlength{\tabcolsep}{12pt} 
\renewcommand{\arraystretch}{1.5} 
\begin{table}[htbp]
\begin{tabular}{c  c } \hline
	\hline
$m_{\rho}$ & 775.2 MeV\\ 
$m_{\pi}$ & 137.5 MeV\\
$m_{K^*}$ & 891.6 MeV \\
$m_{K}$ & 497.6 MeV\\
$m_{\omega}$ & 782.6 MeV\\
$f_{\pi}$ & 93 MeV\\
$\mu$ & 1 GeV \\
\hline
\end{tabular}
\caption{Input parameter values}
\label{sr_parameter}
\end{table}
\end{center}
\endgroup

\section{Result}

In this section, we present results evaluated from the unitarization method. We consider two cases: SU(2) in the chiral limit and SU(3). 

\subsection{SU(2) in the Chiral limit}
Firstly, we consider the flavour SU(2) case in chiral limit, where we take $m_\pi=0$ but take the physical values for $m_\rho$ and $f_\pi$ as given in Table \ref{sr_parameter}. In this case, the resonances are generated from the $\rho \pi - \rho \pi$ channel. Among the possible isospin states ($I = 0,1,2$) in the channel, we are interested only in the isospin triplet states ($I=1$) because it is not clear whether the other isospin states can be directly interpreted as specific meson states. The quantum numbers of this channel coincide with those of the $a_{1}$ resonance($I^{G},J^{P}$) = ($1^-,1^+$).

\begin{figure}[h]
\centerline{
\includegraphics[width=9 cm]{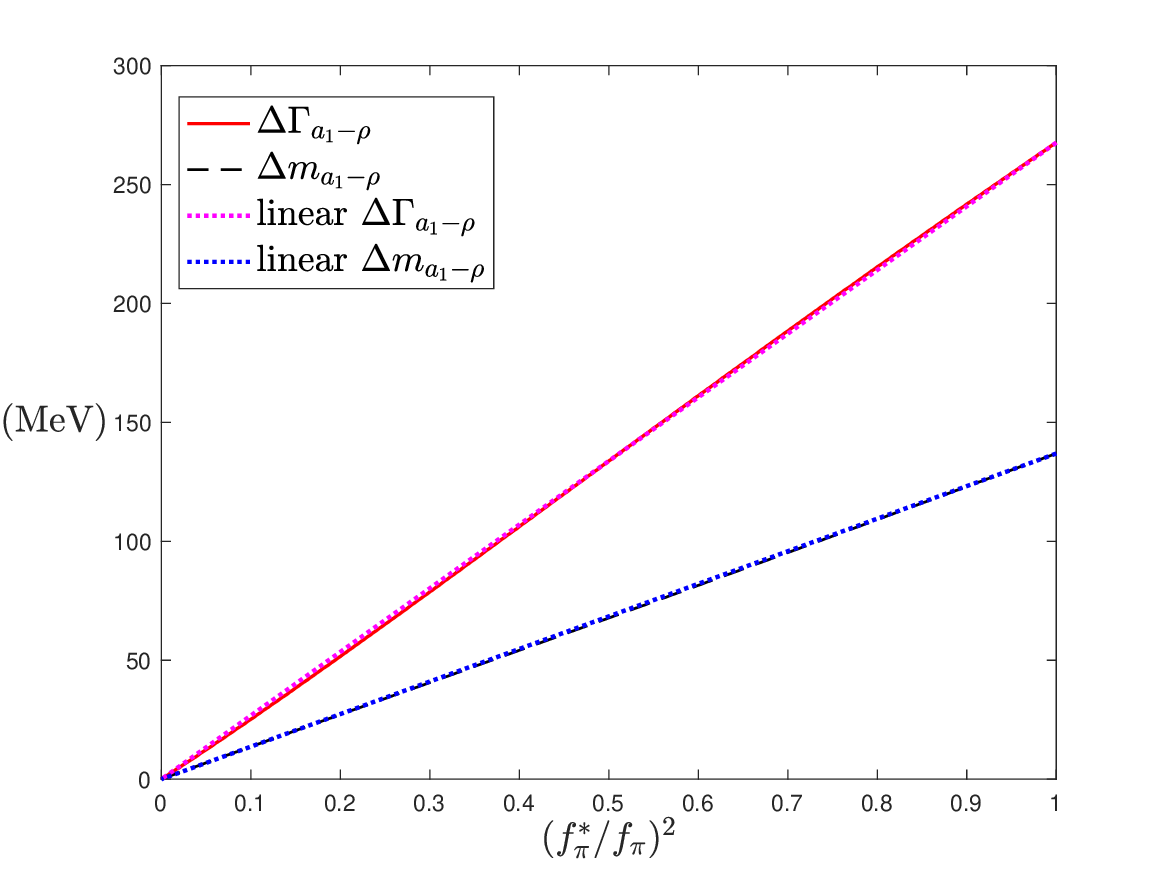}}
\caption{The black dashed line represents the difference between the vacuum mass of $\rho$ meson and the evaluated $a_{1}$ meson mass as a function of the pion decay constant normalized to the vacuum value. The red line represents the decay width of $a_{1}$ meson. The dotted lines stand for the linear fitting in Eq.~\eqref{linear}. The values are calculated in the  limit($m_{\pi} = 0$).}
\label{chiral_dim}
\end{figure}

Figure \ref{chiral_dim} shows the mass difference of the chiral partners and the $a_{1}$ meson decay width, which can be interpreted as the 
decay width difference between the chiral partners($\rho-a_{1}$) since we do not consider the $\rho$ meson decay width. We find that the results can be fitted well as linear functions of the pion decay constant square.
\begin{equation} 
\begin{split} 
m_{a_{1}}-m_{\rho} =&\; 136.9 \bigg(\frac{f_{\pi}^{*}}{f_{\pi}}\bigg)^{2},\\
\Gamma_{a_{1}} =&\; 267.6\bigg(\frac{f_{\pi}^{*}}{f_{\pi}}\bigg)^{2}.
\label{linear}
\end{split} 
\end{equation}

\begin{figure}[h]
\centerline{
\includegraphics[width=9 cm]{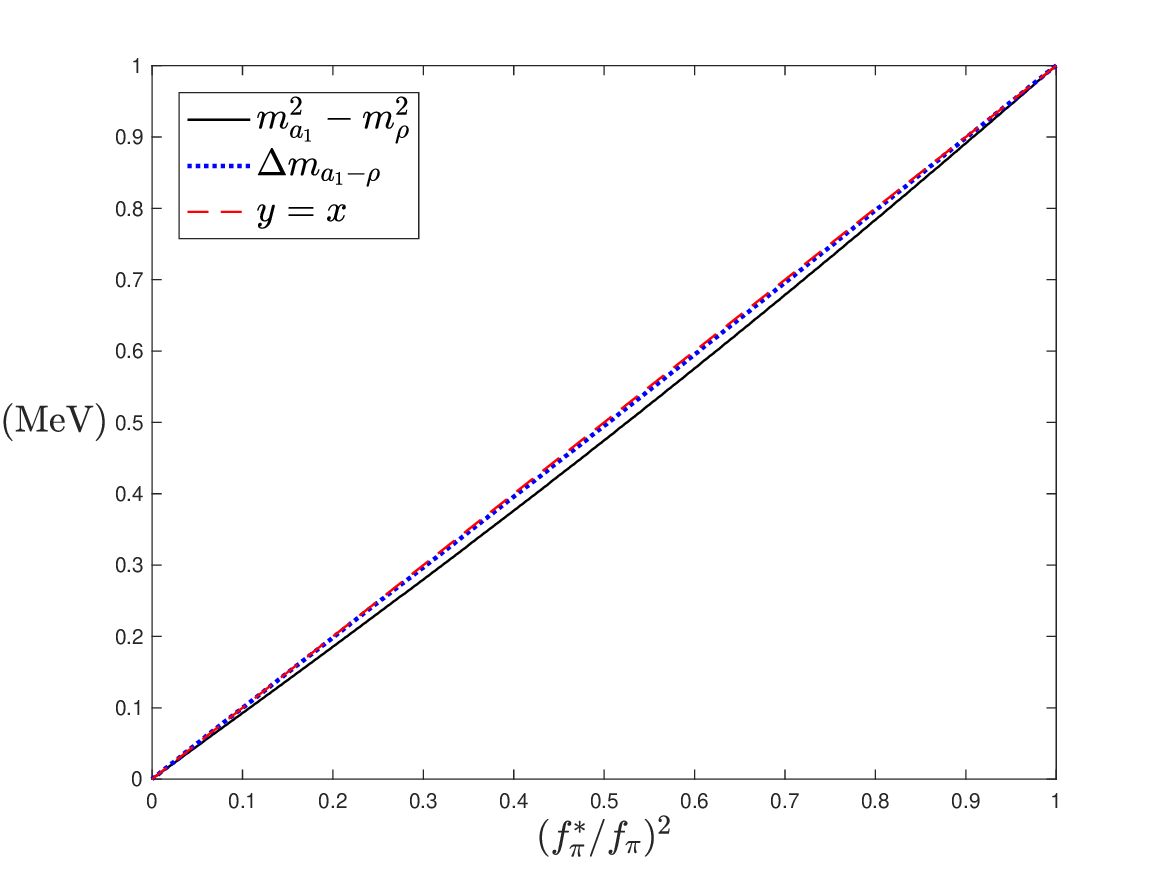}}
\caption{The normalized dependence of the mass square difference in comparison to the mass difference.}
\label{mass2}
\end{figure}

It should be noted that, by generalizing the Weinberg type sum rule, one finds that the difference of mass squares scales as the chiral order parameter\cite{Lee:2023ofg}. As can be seen in Figure \ref{mass2}, this is also consistent with the present result to a lesser degree.

In the unitary approach, as pointed out in Ref.~\cite{Roca:2005nm}, the large uncertainty in the $a_{1}$ decay width may lead to an underestimation of the mass difference. Additionally, both the mass and the decay width values depend on the choice of the subtraction constant $a(\mu)$ in Eq.~\eqref{loopfn}.
Nevertheless, despite these uncertainties, the approximate linear dependence in Eq.~\eqref{linear} remains valid.

\subsection{SU(3) with physical pion mass}
\label{su3}
In this subsection, we investigate two axial vector meson resonances, namely $a_{1}$ and $f_{1}$. Note that isospin singlet $f_{1}$ resonance is a flavor singlet in SU(2) and part of the nonet in SU(3) with symmetry breaking in which the octet is mixed with the singlet. Hence, while $\rho-a_1$ are chiral partners, $\omega-f_1$ are parity parnters\cite{Gubler:2016djf}
To compare the behavior of chiral partners and parity partners, we consider the SU(3) chiral Lagrangian and will use the physical values of the pseudo-Goldstone boson masses. Furthermore, we will use the same value and scaling behavior for the decay constants of all pseudo-Goldstone bosons.

\begin{figure}[h]
\centerline{
\includegraphics[width=9 cm]{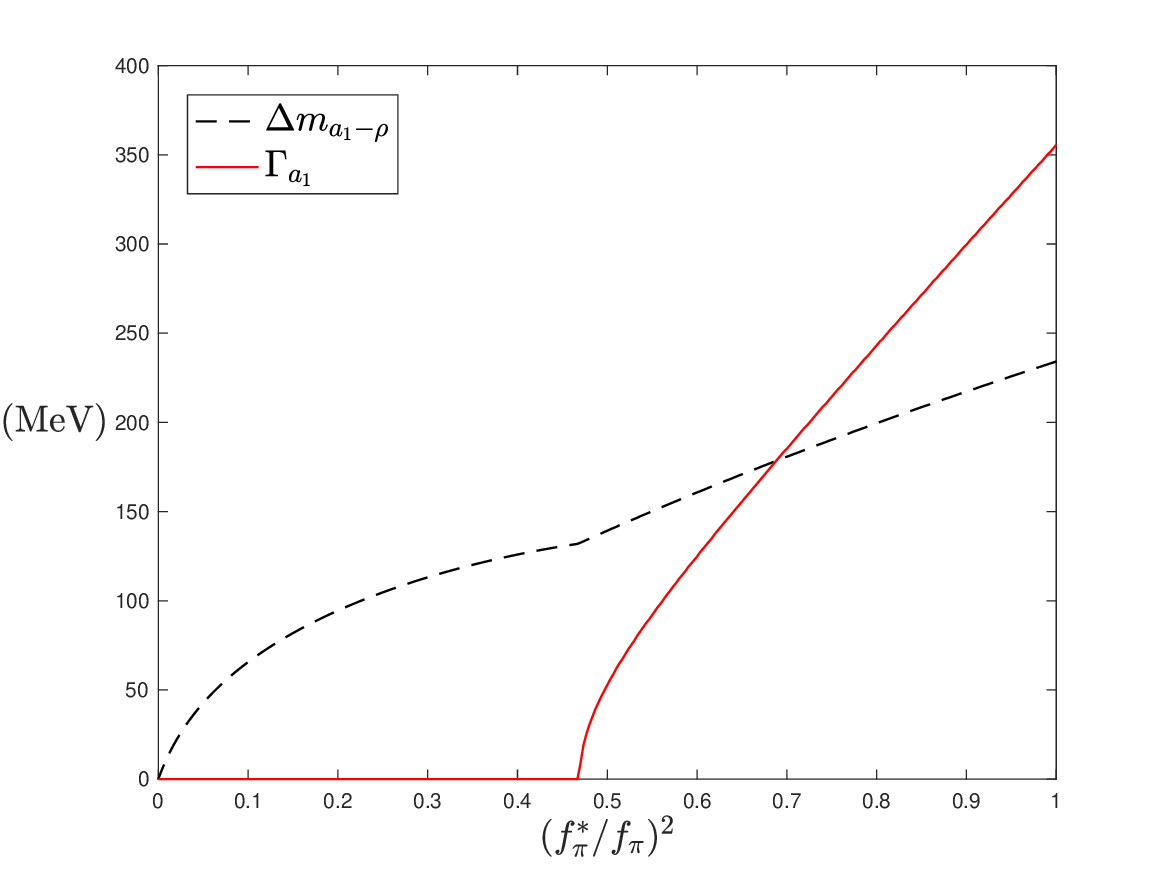}}
\caption{The black dashed line represents the difference between the vacuum mass of the $\rho$ meson and the evaluated mass of the $a_{1}$ meson as a function of the pion decay constant, normalized to the vacuum value. The red line represents the decay width of $a_{1}$ meson. The values are calculated with the physical pion mass($m_{\pi} = 137.5$ MeV).}
\label{a1}
\end{figure}

In Figure~\ref{a1}, we illustrate the shift in the $\rho-a_{1}$ mass difference and the decay width of the $a_{1}$ resonance under symmetry restoration. As previously pointed out in Ref.~\cite{Roca:2005nm}, the $a_{1}$ resonance is predominantly influenced by the $\rho\pi$ channel. Consequently, the results remain largely consistent whether we include the channels containing the $K^{*}K$ state or not.  Unlike the chiral limit, the finite mass of the pion results in a gap between the vector meson mass and the threshold energy of the $\rho-\pi$ channel. 
As shown in Fig~\ref{a1}, the mass difference above the threshold exhibits a linear dependence similar to the previous case. 
Below the threshold, the $a_{1}$ meson properties arise from the bound state of $\rho$ and $\pi$ in this approach, resulting in a different behavior as a function of the chiral order parameter.

\begin{figure}[h]
\centerline{
\includegraphics[width=9 cm]{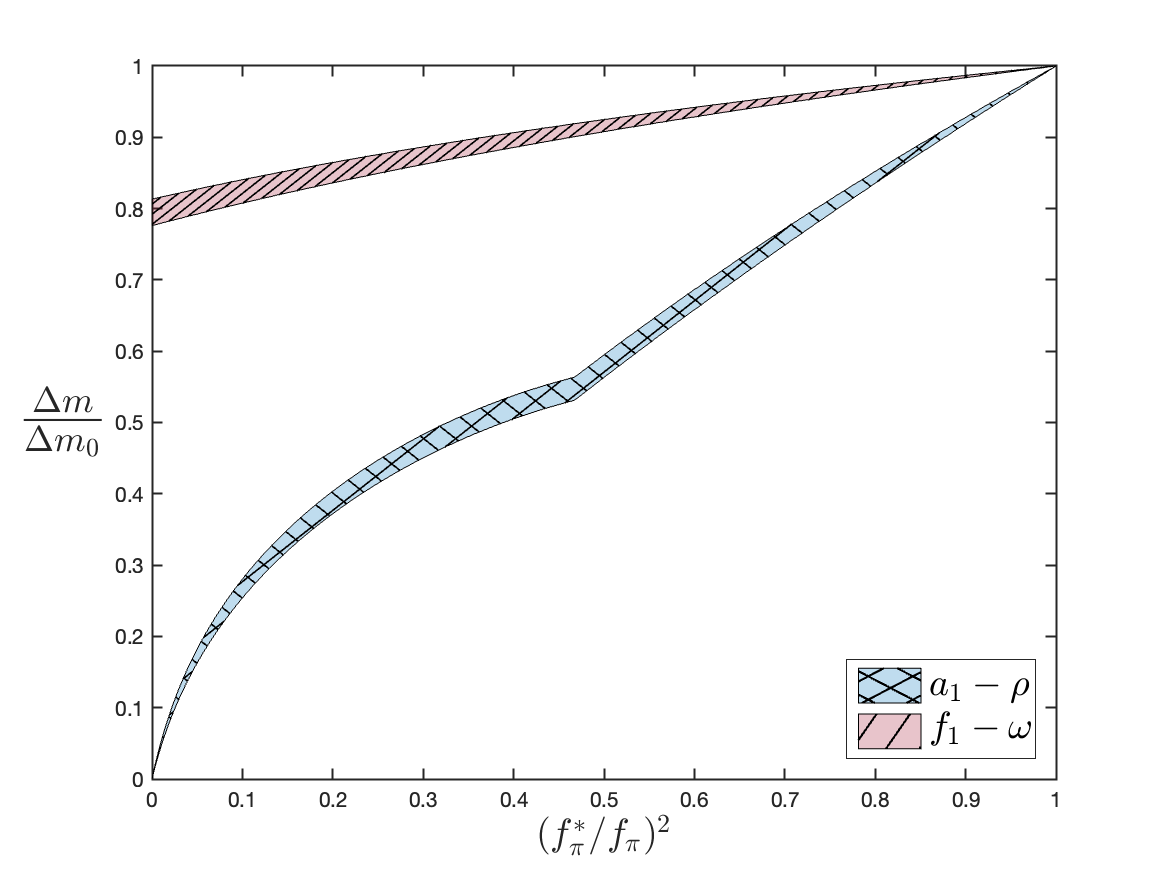}}
\caption{The normalized mass difference between the chiral partners versus the pion decay constant normalized with its vacuum value. The upper bound of each patch represents the mass difference($m_{A}-m_{V}$) normalized as the vacuum value. The lower bound of each patch stands for the mass square difference($m_{A}^{2}-m_{V}^{2}$).}
\label{massdiff}
\end{figure}

Fig~\ref{massdiff} shows that depending on whether or not they form chiral partners, the mass difference behaves differently. The chiral partners, $\rho$ and $a_{1}$ have degenerate mass states when chiral symmetry is restored. Meanwhile, the significant fraction of the mass difference between the parity partners remains even after the symmetry restoration\cite{Gubler:2016djf,Kim:2021xyp}. 

\section{conclusions}
In this study, we have explored the modifications to the properties of axial-vector mesons associated with chiral symmetry restoration. By manipulating the value of $f_{\pi}^{2}$ in the potential $V(s)$, we have investigated how these properties evolve during the restoration of chiral symmetry. In the chiral limit, both the mass and decay width of the $a_{1}$ meson exhibits a linear decrease as the order parameter decreases. However, in the presence of explicit symmetry breaking, these properties do not change linearly but retain a monotonic decreasing behavior.

We conclude that the mass difference between chiral partners vanishes when chiral symmetry is restored. On the other hand, the mass of $f_{1}$ barely decreases and the mass difference to its parity partner remains large even when chiral symmetry is restored. This implies that the spontaneous breakdown of chiral symmetry has a minor effect on the generation of the mass difference between parity partners, consistent with the QCD sum rule result and expected from general consideration\cite{Gubler:2016djf}.

\section*{Acknowledgements}
This work was supported by Samsung Science and Technology
Foundation under Project Number SSTF-BA1901-04, and by the Korea National Research Foundation under grant number No. 2023R1A2C3003023.

%_______________________________________ref_______________________________%

\end{document}